\documentclass[superscriptaddress,amsmath,amssymb,aps,prd,notitlepage,longbibliography,floatfix,nofootinbib]{revtex4-1}

\usepackage{tensor}     
\usepackage{graphicx}   
\usepackage{subfigure}
\usepackage[
colorlinks=true,        
citecolor=blue,         
linkcolor=blue,         
urlcolor=blue           
]{hyperref}             
\usepackage{bm}         
\usepackage{xcolor}     
\usepackage{lipsum}
\usepackage{color}      
\usepackage[utf8]{inputenc} 
\usepackage[section]{placeins} 
\usepackage{multirow}
\usepackage[title]{appendix}
\allowdisplaybreaks[4]
\newcommand{\nc}{\newcommand*}

\nc{\al}{\alpha}
\nc{\s}{\sigma}
\nc{\kp}{\kappa}
\nc{\dt}{\delta}
\nc{\Dt}{\Delta}
\nc{\Ld}{\Lambda}
\nc{\p}{\partial}
\nc{\Gm}{\Gamma}
\nc{\om}{\omega}
\nc{\Om}{\Omega}
\nc{\rd}{\mathrm{d}}
\def\({\left(}
\def\){\right)}
\def\[{\left[}
\def\]{\right]}
\def\e{\begin{equation}}
\def\q{\end{equation}}
\def\m{\begin{eqnarray}}
\def\n{\end{eqnarray}}
\nc{\Eq}[1]{Eq.~\eqref{#1}}     
\nc{\Fig}[1]{Fig.~\ref{#1}}     
\nc{\Table}[1]{Table~\ref{#1}}  
\nc{\Sec}[1]{Sec.~\ref{#1}}     
\nc{\Msun}{M_\odot}             
\nc{\fpbh}{f_{\mathrm{pbh}}}    
\nc{\fpbhn}{f_{\mathrm{pbh0}}}    
\nc{\mR}{\mathcal{R}} 
\nc{\seq}{\sigma_{\mathrm{eq}}}
\nc{\ogw}{\Omega_{\mathrm{GW}}}
\nc{\gpcyr}{\mathrm{Gpc}^{-3}\,\mathrm{yr}^{-1}}
\nc{\lvc}{LIGO/Virgo} 
\nc{\SNR}{\mathrm{SNR}} 
\nc{\mmin}{{m_{\mathrm{min}}}}
\nc{\mmax}{{m_{\mathrm{max}}}}
\nc{\Mmin}{{M_{\mathrm{min}}}}
\nc{\fmin}{{f_{\mathrm{min}}}}
\nc{\VT}{\mathrm{VT}}
\nc{\rhoGW}{\rho_{\mathrm{GW}}}
\nc{\vth}{\vec{\theta}}
\nc{\vd}{\vec{d}}
\nc{\vla}{\vec{\lambda}}
\nc{\Nobs}{N_{\mathrm{obs}}}
\nc{\av}[1]{\langle #1 \rangle} 
\nc{\km}{\mathrm{km}}
\nc{\Mpc}{\mathrm{Mpc}}
\nc{\Tobs}{T_{\mathrm{obs}}}
\nc{\Ntemp}{N_{\mathrm{temp}}}
\nc{\fyr}{f_{\mathrm{yr}}}
\nc{\addref}{[\textcolor{red}{add ref}] } 
\nc{\eg}{\textit{e.g.~}}
\nc{\app}{\approx}
\nc{\hf}{\frac{1}{2}}
\nc{\discuss}{\textcolor{red}{Add discussion here!}}
\nc{\red}[1]{\textcolor{red}{#1}}
\nc{\hp}{h_+} 
\nc{\hc}{h_{\times}} 
\nc{\Oh}{\hat{\Omega}}
\nc{\vx}{\vec{x}}
\nc{\mh}{\hat{m}}
\nc{\nh}{\hat{n}}
\nc{\zh}{\hat{z}}
\nc{\ph}{\hat{p}}
\nc{\A}[1]{\mathcal{A}_{#1}}
\nc{\bn}[1]{\dt\bm{t}_{\text{#1}}}
\nc{\bC}[1]{\bm{C}_{\text{#1}}}
\nc{\NTOA}{N_{\text{TOA}}}
\nc{\Nmode}{{N_{\text{mode}}}}
\nc{\ARN}{A_{\rm{RN}}}
\nc{\gRN}{\gamma_{\rm{RN}}}
\nc{\bS}{\mathbf{\Sigma}}
\nc{\br}{\mathbf{r}}
\nc{\bN}{\mathbf{R}}
\nc{\Agw}{A_\mathrm{GWB}}
\nc{\UCP}{\mathrm{UCP}}
\nc{\TT}{\mathrm{TT}}
\nc{\ST}{\mathrm{ST}}
\nc{\SL}{\mathrm{SL}}
\nc{\VL}{\mathrm{VL}}

\nc{\BFST}{$107 \pm 7$}

\begin{document}
	
\title{Generic Modified Teukolsky Formalism beyond General Relativity for Spherically Symmetric Cases}

	

\author{Rong-Zhen Guo}
\email{guorongzhen@ucas.ac.cn}
\affiliation{School of Fundamental Physics and Mathematical Sciences
Hangzhou Institute for Advanced Study, UCAS, Hangzhou 310024, China}
\affiliation{School of Physical Sciences,
	University of Chinese Academy of Sciences,
	No. 19A Yuquan Road, Beijing 100049, China}

\author{Hongwei Tan}
\email{htan2018@fau.edu}
\affiliation{Department of Astronomy, School of Physics and Materials Science, Guangzhou
University, Guangzhou 510006,China}

\author{Qing-Guo Huang}
\email{Corresponding author: huangqg@itp.ac.cn}
\affiliation{School of Fundamental Physics and Mathematical Sciences
Hangzhou Institute for Advanced Study, UCAS, Hangzhou 310024, China}
\affiliation{School of Physical Sciences,
University of Chinese Academy of Sciences,
No. 19A Yuquan Road, Beijing 100049, China}
\affiliation{CAS Key Laboratory of Theoretical Physics,
Institute of Theoretical Physics, Chinese Academy of Sciences,
Beijing 100190, China}

\date{\today}

\begin{abstract}
The observation of gravitational waves has inaugurated a new era for testing gravitational theories in strong-field, nonlinear regimes. Gravitational waves emit during the ringdown phase of binary black hole mergers and from extreme mass ratio inspirals are particularly sensitive to the properties of black holes, making them crucial for probing deviations from general relativity. These studies 
need a robust foundation in black hole perturbation theory beyond general relativity. While existing studies have employed black hole perturbation theories to explore modifications beyond general relativity, they often focus on specific alternative theories or phenomenological models of quantum gravity. In this paper, we establish a modified decoupled Teukolsky formalism that is broadly applicable to spherically symmetric spacetimes without requiring a predetermined gravitational Lagrangian. This formalism uses the Newman-Penrose framework, which utilizes curvature perturbations characterized by Weyl scalars, to accommodate a wider class of spacetimes beyond general relativity. Our approach correctly handles non-Ricci-flat backgrounds and circumvents subtle analytical issues associated with effective potentials that are present in other modified Teukolsky formalisms. 
\end{abstract}

\maketitle
\maketitle
\section{Introduction}
 The discovery of gravitational waves (GWs) has ushered in a new era for testing gravitational theory in strong-field, nonlinear regimes \cite{TheLIGOScientific:2014jea,TheVirgo:2014hva,TheLIGOScientific:2016htt,TheLIGOScientific:2016src,O1_release,Will:2014kxa}.  The next generation of gravitational wave detectors, whether space-based or ground-based, will enable stronger tests of gravitational theory, especially the deviations from 
the predictions of Einstein’s general relativity (GR),  the most successful theory of gravity to date \cite{Berti:2018cxi,Reitze:2019iox,Baker:2019nia,Perkins:2020tra,TianQin:2015yph,Hu:2017mde}. 

In order to probe gravity in strong-field regions, GWs from the ringdown phase of binary black hole (BH) mergers and extreme mass ratio inspirals (EMRIs) are important. Ringdown phase is the final stage of the coalescence of binary BH systems. After the formation of final BHs, GWs from ringdown phase are characterized by the superposition of numerous quasinormal
modes (QNMs), each with a complex eigenfrequency  \cite{Berti:2015itd,Berti:2009kk}. EMRI is a binary system composed of stellar-mass compact objects, inspiralling into massive BHs at the center of galaxies, where the mass ratio $q$ between two objects is $10^{-8} \lesssim q \lesssim 10^{-5}$ \cite{LISA:2022yao,Amaro-Seoane:2012lgq}. Since GWs generated by these two sources are sensitive to the properties of 
BHs, many studies have explored the impact of modifications beyond GR. In \cite{Berti:2004md,Liu:2020ola,Flachi:2012nv,Konoplya:2019xmn}, authors investigate the impact of quantum gravity on QNMs. GW echoes, which can also be produced during ringdown stage, are examined  in \cite{Wang:2019rcf,Maggio:2019zyv,Ikeda:2021uvc,Dey:2020lhq}. The deviation towards GR in EMRIs are analyzed in \cite{Maggio:2021uge,Fang:2021iyf,Cardoso:2018zhm,Pani:2011xj}.

The studies mentioned above are based on BH perturbation theories. In GR, there are two well-established formalisms within BH perturbation theory. The first type directly perturbs the background metric, resulting either in the perturbation equations for different metric components \cite{Clarkson:2002jz,Sarbach:2001qq} or in the construction of master equations such as  Regge-Wheeler-Zerilli equations \cite{Regge:1957td,Zerilli:1970se}. The second type uses curvature perturbations, which is first calculated on Kerr spacetime by S. Teukolsky in \cite{Teukolsky:1973ha,Press:1973zz,Teukolsky:1974yv}. It is based on the Newman-Penrose (NP) formalism \cite{Newman:1961qr}.  Curvature perturbations in NP formalism is characterized by quantities called Weyl scalars. The equations for the curvature perturbations 
corresponding to incoming and outgoing GWs, i.e., $\Psi_0$ and $\Psi_4$, would be decoupled and separable in the background of spinning BHs in GR. Both methods are effective in GR and can be transformed into each other through Chandrasekhar-Sasaki-Nakamura transformation \cite{Sasaki:2003xr}, thus can support consistent physical predictions.

There are various studies on BH perturbation theories beyond GR. Some of these studies focus on specific modified gravity theories. For example, the authors studied perturbations of spinning BHs in dynamical Chern-Simons gravity \cite{Yunes:2009hc} using NP formalism in \cite{Wagle:2023fwl}. Direct calculation towards metric perturbations within the same theory is established in \cite{Cardoso:2009pk}. The Einstein-dilaton-Gauss-Bonnet gravity is studied in \cite{Blazquez-Salcedo:2016enn,Pierini:2021jxd}. 
For other BH perturbation theories in modofied gravity, such as Horndeski gravity, Einstein-Aether theory, effective-field-theory based gravity, see\cite{Tattersall:2018nve,Konoplya:2006rv,deRham:2020ejn}. There are also studies focus on BHs that takes account of quantum effects. Perturbative phenomena on BH models of loop quantum gravity (LQG) is studied in \cite{Shao:2023qlt,Yang:2022btw}. GWs from BH area quantization or other micro-structures of event horizons are calculated in \cite{Coates:2021dlg,Cardoso:2019apo,Guo:2021xao,Guo:2023mel}. In recent years, there have been attempts to establish a more universal BH perturbation theory, which could cover as many scenarios beyond GR as possible. In \cite{Li:2022pcy,Hussain:2022ins}, the authors develope the modified Teukolsky formalism to study non-Ricci-flat, Petrov Type I BHs. In the study of BH Hawking radiation, a similar 
modified Teukolsky formalism has been used, applicable to general spherically symmetric spacetimes \cite{Arbey:2021jif,Arbey:2021yke}.

In this work, we establish a modified decoupled Teukolsky formalism that is broadly suitable for spherically symmetric spacetimes, regardless of whether a predetermined Lagrangian is specified, thus could cover BHs based on whether classical modified gravity or phenomenological models of quantum gravity. In addition to its broad applicability, our formalism also possesses some advantages compared to other modified Teukolsky formalisms. Our equations correctly 
treat the non-Ricci-flatness of the background and avoid the presence of subtle analytical properties of effective potentials.

The remainder of this paper is organized as follows. In Sec.\ref{Sec1}, we present a brief introduction to the NP formalism and relevant NP equations, which is useful to derive Teukolsky's equations in GR. In Sec.\ref{Sec2}, we provide our modified Teukolsky formalism. We discuss the advantages of NP formalism, which can be a solid foundation of BH perturbation theories beyond GR. Subsequently, we derive the decoupled Teukolsky's equation in spherically symmetric spacetimes that satisfy the so-called $tr$ symmetry using gauge fixing method in NP formalism. The cases without $tr$ symmetry are also discussed. In Sec.\ref{Sec3}, we compare our 
equations with that of previous researchers. Finally, in Sec.\ref{Sec4}, we summarize our work and discuss some future avenues. Henceforth, we  work in 4-dimensional spacetimes with metric signature $(+,-,-,-)$, which is consistent with other literatures of NP formalism \cite{Newman:1961qr}. We use the units $c=G=\hbar=1$.

\section{Quick reviews of NP formalism in GR}\label{Sec1}
NP formalism is a tetrad formalism with the choice of null basis vectors\cite{Chandrasekhar:1985kt}, which was first introduced by Ezra Newman and Roger Penrose in 1962 \cite{Newman:1961qr}. Using this framework, Teukolsky developed a formalism to study the dynamical perturbation of spinning BHs in GR \cite{Teukolsky:1973ha,Press:1973zz,Teukolsky:1974yv}. In this section we take a brief introduction to relevant equations, which are directly related to the decoupled perturbation equations in NP formalism. Readers familiar with this topic can skip to the next section.

\subsection{NP Formalism}\label{review}

In NP formalism, we choose a pair of real-valued null vectors $l^\mu, n^\mu$ and a pair of comlpex-valued null vectors $m^\mu, \bar{m}^\mu$ that are complex conjugates of each other as tetrad basis at every point of a four-dimensional pseudo-Riemannian manifold with signature $-2$ and metric $g_{\mu \nu}$, which satisfy
\begin{equation}
\begin{aligned}
l_\mu l^\mu=n_\mu n^\mu & =m_\mu m^\mu=\bar{m}_\mu \bar{m}^\mu=0, \\
l_\mu n^\mu & =-m_\mu \bar{m}^\mu=1, \\
l_\mu m^\mu=l_\mu \bar{m}^\mu & =n_\mu m^\mu=n_\mu \bar{m}^\mu=0,
\label{OrthCond}
\end{aligned}
\end{equation}
where $\bar{\ }$ denotes the complex conjugate. So the metric $g_{\mu \nu}$ can be also expressed as
\begin{equation}
g_{\mu \nu}=l_\mu n_\nu+n_\mu l_\nu-m_\mu \bar{m}_\nu-\bar{m}_\mu m_\nu.
\end{equation}
The directional derivatives are designed by
\begin{equation}
D  \equiv  l^\mu \nabla_{\mu},  \Delta  \equiv  n^\mu \nabla_{\mu} ,\delta 
\equiv  m^\mu \nabla_{\mu},  \bar{\delta}\equiv  \bar{m}^\mu \nabla_{\mu}.
\end{equation}
Here $\nabla_\mu$ is compatible with the spacetime metric $g_{\mu\nu}$, $\nabla_\mu g_{\nu\sigma}=0$.

In this formalism, fundamental variables contain 12 spin coffeicients $\{\kappa,\sigma,\gamma,\nu,\rho,\mu,\tau,\pi,\epsilon,\gamma,\alpha,\beta\}$\footnote{There are numerous symbols in NP formalism, with $\pi$ denoting a specific NP scalar. So we use $\varpi$ to represent mathematical constant $\pi$, thereby mitigating potential notational confusion.}, corresponding to Ricci rotation coefficients \cite{Wald:1984rg}, 5 Weyl scalars $\{\Psi_i\}_{i=0,...,4}$ and 10 Ricci scalars $\{\Phi_{ij},\Lambda\}_{i,j=0,1,2}$, corresponding to independent parts of Riemann tensor. Explicit definition can be checked in Appendix\ref{NP}. One can use these quantities to establish fundamental relations in 
NP formalism, such as Ricci identities and Bianchi identities \cite{Newman:1961qr}. The following equations are the materials needed for deriving the Teukolsky's equation in GR, listed as follows:
\begin{equation}
D \Psi_1-\bar{\delta} \Psi_0-D \Phi_{01}+  \delta \Phi_{00}  =(\pi-4 \alpha) \Psi_0+2(2 \rho+\epsilon) \Psi_1-3 \kappa \Psi_2  -(\bar{\pi}-2 \bar{\alpha}-2 \beta) \Phi_{00}-2(\bar{\rho}+\epsilon) \Phi_{01}  -2 \sigma \Phi_{10}+2 \kappa \Phi_{11}+\bar{\kappa} \Phi_{02},\label{Teukolsky1}
\end{equation}

\begin{equation}
\Delta \Psi_0-\delta \Psi_1+D \Phi_{02}-  \delta \Phi_{01}  =(4 \gamma-\mu) \Psi_0-2(2 \tau+\beta) \Psi_1+3 \sigma \Psi_2  -\bar{\lambda} \Phi_{00}+2(\bar{\pi}-\beta) \Phi_{01}+2 \sigma \Phi_{11}  +(\bar{\rho}+2 \epsilon-2 \bar{\epsilon}) \Phi_{02}-2 \kappa \Phi_{12},\label{Teukolsky2}
\end{equation}

\begin{equation}
\Delta \Psi_3-\delta \Psi_4-\Delta \Phi_{21}+  \bar{\delta} \Phi_{22}  =3 \nu \Psi_2-2(\gamma+2 \mu) \Psi_3+(4 \beta-\tau) \Psi_4  -2 \nu \Phi_{11}-\bar{\nu} \Phi_{20}+2 \lambda \Phi_{12}  +2(\gamma+\bar{\mu}) \Phi_{21}+(\bar{\tau}-2 \bar{\beta}-2 \alpha) \Phi_{22},\label{Teukolsky11}
\end{equation}

\begin{equation}
D \Psi_4-\bar{\delta} \Psi_3+\Delta \Phi_{20}-  \bar{\delta} \Phi_{21}  =-3 \lambda \Psi_2+2(\alpha+2 \pi) \Psi_3+(\rho-4 \epsilon) \Psi_4
 +2 \nu \Phi_{10}-2 \lambda \Phi_{11}-(2 \gamma-2 \bar{\gamma}+\bar{\mu}) \Phi_{20}  -2(\bar{\tau}-\alpha) \Phi_{21}+\bar{\sigma} \Phi_{22},\label{Teukolsky22}
\end{equation}
which are Bianchi identities, and
\begin{equation}
D \sigma-\delta \kappa=(\rho+\bar{\rho}+3 \epsilon-\bar{\epsilon}) \sigma-(\tau-\bar{\pi}+\bar{\alpha}+3 \beta) \kappa+\Psi_0,\label{Teukolsky3}
\end{equation}
\begin{equation}
\Delta \lambda-\bar{\delta} \nu=-(\mu+\bar{\mu}+3 \gamma-\bar{\gamma}) \lambda+(3 \alpha+\bar{\beta}+\pi-\bar{\tau}) \nu-\Psi_4,\label{Teukolsky33}
\end{equation}
which belong to  Ricci identities.
\subsection{Teukolsky's equation}
In NP formalism, one can classify spacetime by the algebraic structure of the Weyl curvature tensor. That is, for any given spacetime, one can always choose an NP tetrad such that the values of its Weyl scalars belong to one of the following categories:
\begin{itemize}
   \item Type I: $\Psi_0=\Psi_4=0$,
   \item Type II: $\Psi_0=\Psi_1=\Psi_4=0$,
   \item Type D: $\Psi_0=\Psi_1=\Psi_3=\Psi_4=0$,
   \item Type III: $\Psi_0=\Psi_1=\Psi_2=\Psi_4=0$,
   \item Type N: $\Psi_0=\Psi_1=\Psi_2=\Psi_3=0$.
\end{itemize}
We call it Petrov classification \cite{Chandrasekhar:1985kt,Petrov:2000bs}. In GR, the most notable BHs are Petrov Type D vacuum solutions, such as the Schwarzschild and Kerr solutions. Thus, if we expand all NP scalars into a background plus a perturbation on Type D spacetime, e.g., $\Psi_i=\Psi_i^{(0)}+\epsilon \Psi_i^{(1)}$ for Weyl scalars, where the superscript (0) represent background quantities, while the superscript (1) stands for  perturbation from this background with $\epsilon$ a book-keeping parameter. One can choose suitable NP tetrad such that
\begin{equation}
\begin{aligned}
\Psi_0^{(0)} & =\Psi_1^{(0)}=\Psi_3^{(0)}=\Psi_4^{(0)}=0, \\
\kappa^{(0)} & =\sigma^{(0)}=\nu^{(0)}=\lambda^{(0)}=0.
\end{aligned}
\label{Sachs}
\end{equation}
according to Goldberg–Sachs theorem \cite{Chandrasekhar:1985kt}.

We are interested in $\Psi_0^{(1)}$ and $\Psi_4^{(1)}$. Indeed, one can check these quantities are gauge invariants at the level of first-order perturbation based on transformation rules in Appendix.\ref{NP}. Besides, according to peeling theorems \cite{Kroon:2016ink}, $\Psi_0$ and $\Psi_4$ can be explained as ingoing and outgoing radiation of a gravitationally isolated system, i.e., asymptotically flat spacetime. Thus, we can expand eqs.(\ref{Teukolsky1},\ref{Teukolsky2},\ref{Teukolsky3}) in the same manner. One can get series of equations related to $\Psi_0{}^{(1)}$:
\begin{equation}
\begin{aligned}
&\left(\bar{\delta}-4 \alpha+\pi\right)^{(0)} \Psi_0{ }^{(1)}-(D-4 \rho-2 \epsilon)^{(0)} \Psi_1{ }^{(1)}-3 \kappa^{(1)} \Psi_2{ }^{(0)}=S_1,  \\
&(\Delta-4 \gamma+\mu)^{(0)} \Psi_0{ }^{(1)}-(\delta-4 \tau-2 \beta)^{(0)} \Psi_1{ }^{(1)}-3 \sigma^{(1)} \Psi_2{ }^{(0)}=S_2,\\
&\left(D-\rho-\bar{\rho}-3 \epsilon+\bar{\epsilon}\right)^{(0)} \sigma^{(1)}-\left(\delta-\tau+\bar{\pi}-\bar{\alpha}-3 \beta\right)^{(0)} \kappa^{(1)}-\Psi_0{ }^{(1)}=0.
\end{aligned}
\label{psi0eq}
\end{equation}
where $S_1,S_2$ are source terms related to the perturbation of Ricci scalars. In GR, Ricci scalars can be directly calculated from stress tensor $T_{\mu\nu}$ by trace-reversed form of Einstein's equation
\begin{equation}
R^{\mu \nu}=8 \varpi\left(T^{\mu \nu}-\frac{1}{2} T g^{\mu \nu}\right).
\end{equation}

In order to eliminate $\kappa_1^{(1)},\sigma_1^{(1)}$ in eq.\eqref{psi0eq}, we can multiply $\Psi_2{ }^{(0)}$ to the third line and plug it into 0-th order operators. We can get
\begin{equation}
    \left(D-\rho-\bar{\rho}-3 \epsilon+\bar{\epsilon}\right) (\Psi_2{ }\sigma^{(1)})-\left(\delta-\tau+\bar{\pi}-\bar{\alpha}-3 \beta\right)( \Psi_2{ }\kappa^{(1)})-\Psi_2\Psi_0{ }^{(1)}-\sigma^{(1)}D \Psi_2+\kappa^{(1)}\delta \Psi_2{ }=0.\label{psi2medium}
\end{equation}
Notice that the labels (0) are dropped from all unperturbed quantities in order to simplify the notations. Acoording to eq.(\ref{Dpsi2}) and eq.(\ref{deltapsi2}), we get
\begin{equation}
D \Psi_2=3 \rho \Psi_2, \quad \delta \Psi_2=3 \tau \Psi_2.
\label{Dpsi22}
\end{equation}
So we can simplify eq.(\ref{psi2medium}) into
\begin{equation}
\left(D-3 \epsilon+\bar{\epsilon}-4 \rho-\bar{\rho}\right) \Psi_2 \sigma^{(1)}-\left(\delta+\bar{\pi}-\bar{\alpha}-3 \beta-4 \tau\right) \Psi_2 \kappa^{(1)}-\Psi_0{}^{(1)} \Psi_2=0.\label{elieq}
\end{equation}
Appling $\left(\delta + \bar{\pi} - \bar{\alpha} - 3 \beta - 4 \tau\right)$ and $\left(D - 3 \epsilon + \bar{\epsilon} - 4 \rho - \bar{\rho}\right)$  to the first and second lines of eq.(\ref{psi0eq}) respectively, the perturbative spin coefficients will cancel out due to eq.(\ref{elieq}). The next step is to eliminate $\Psi_{1}{}^{(1)}$ term. And there exists an background operator commutation relation
\begin{equation}
\left[D-(p+1) \epsilon+\bar{\epsilon}+q \rho-\bar{\rho}\right](\delta-p \beta+q \tau)  -\left[\delta-(p+1) \beta-\bar{\alpha}+\bar{\pi}+q \tau\right](D-p \epsilon+q \rho)=0,
\label{comimportant}
\end{equation}
which always holds for vacuum spacetimes of Type D. Here $p,q$ are arbitrary constants. This relation can be proved by eqs.({\ref{comrel}},\ref{NPuse2},\ref{NPuse3},\ref{NPuse1}) \cite{Teukolsky:1973ha}. Thus, $\Psi_{1}{}^{(1)}$ term then vanish by eq.(\ref{comimportant}) with $p = 2, q = -4$. The final decoupled equation for $\Psi_0^{(1)}$ is the one of so called Teukolsky's equations in BH perturbation theory:
\begin{equation}
\left[(D-3 \epsilon+\bar{\epsilon}-4 \rho-\bar{\rho})(\Delta-4 \gamma+\mu)-(\delta+\bar{\pi}-\bar{\alpha}-3 \beta-4 \tau)(\bar{\delta}+\pi-4 \alpha)-3 \Psi_2\right] \Psi_0{}^{(1)}=4 \varpi T_0,
\end{equation} 
where
\begin{equation}
\begin{aligned}
T_0 & =(\delta+\bar{\pi}-\bar{\alpha}-3 \beta-4 \tau)\left[(D-2 \epsilon-2 \bar{\rho}) T_{l m}-(\delta+\bar{\pi}-2 \bar{\alpha}-2 \beta) T_{l l}\right] \\
& +(D-3 \epsilon+\bar{\epsilon}-4 \rho-\bar{\rho})\left[(\delta+2 \bar{\pi}-2 \beta) T_{l m}-(D-2 \epsilon+2 \bar{\epsilon}-\bar{\rho}) T_{m m}\right].
\end{aligned}
\end{equation}

The above procedure can be also similarly applied to $\Psi_4^{(1)}$ based on eqs.(\ref{Teukolsky11},\ref{Teukolsky22},\ref{Teukolsky33}). However, a more convenient method is to perform the so-called prime operation \cite{ODonnell:2003lqh,Penrose:1985bww}:
\begin{equation}
l^{\mu}  \mapsto n^{\mu} ,
n^{\mu}  \mapsto l^{\mu} ,
m^{\mu}  \mapsto \bar{m}^{\mu},
\bar{m}^{\mu} \mapsto m^{\mu}.
\end{equation}
The full set of NP equations is invariant under this operation, which can be checked by relations in Appendix.\ref{NP}. Thus, Teukolsky's equation for $\Psi_4^{(1)}$ is
\begin{equation}
\left[(\Delta+3 \gamma-\bar{\gamma}+4 \mu+\bar{\mu})(D+4 \epsilon-\rho)-(\bar{\delta}-\bar{\tau}+\bar{\beta}+3 \alpha+4 \pi)(\delta-\tau+4 \beta)-3 \Psi_2\right] \Psi_4^{(1)}=4 \varpi T_4,
\end{equation}
where 
\begin{equation}
\begin{aligned}
T_4 & =(\Delta+3 \gamma-\bar{\gamma}+4 \mu+\bar{\mu})\left[(\bar{\delta}-2 \bar{\tau}+2 \alpha) T_{n \bar{m}}-(\Delta+2 \gamma-2 \bar{\gamma}+\bar{\mu}) T_{\bar{m} \bar{m}}\right] \\
& +(\bar{\delta}-\bar{\tau}+\bar{\beta}+3 \alpha+4 \pi)\left[(\Delta+2 \gamma+2 \bar{\mu}) T_{n \bar{m}}-(\bar{\delta}-\bar{\tau}+2 \bar{\beta}+2 \alpha) T_{n n}\right].
\end{aligned}
\end{equation}

\section{Modified Teukolsky's equations beyond GR}\label{Sec2}
This section addresses the decoupled gauge-invariant curvature perturbation equations in arbitrary spherically symmetric asymptotically flat spacetimes beyond GR. We first highlight the advantages of the NP formalism as a perturbative method. Next, we calculate the modified Teukolsky's equations for spherically symmetric spacetimes with so-called “tr” symmetry and discuss methods for more general cases. Finally, we compare our results with state-of-the-art equations \cite{Li:2022pcy,Arbey:2021jif}, elucidating the strengths and limitations of various approaches.

\subsection{Advantages of the NP formalism}
In spherically symmetric spacetimes, perturbation equations can be categorized into three types: coupled partial differential equations for metric components, the Regge-Wheeler-Zerilli type master equations, and the Teukolsky's equations (also known as the Bardeen-Press equation in spherical symmetry spacetime). All of these methods have been widely used in BH perturbation theory, including the calculation of QNM, tidal deformation, and self force. Teukolsky's equation is the least straightforward one to deal with perturbative problems, since it deals with the curvature perturbations, not the metric perturbations themselves. This also makes the metric reconstruction using the solution of the Teukolsky's equation is  delicate even in GR \cite{Loutrel:2020wbw,Whiting:2005hr,Ori:2002uv}. 

Even so, the Teukolsky's equation has a formal simplicity that the other two methods do not have. From the derivation of Teukolsky's equations, one will notice that all the NP equations used above are geometric equations on a pseudo-Riemannian geometry, rather than the equations of motion themselves, i.e., Einstein equations. Indeed, the equations of motion are only considered if one wants to establish the relations between the Ricci scalars and the components of stress tensor. Therefore, as long as the gravitational theory being studied includes the metric as a variable and this metric field is defined within the framework of pseudo-Riemannian geometry, then if the background spacetime remains of Petrov type D, the calculation procedures in 
GR can be extended to more general cases, though the effects of non-zero Ricci scalars and their first-order perturbations must be considered. And the Ricci scalars contributed by the background are particularly noteworthy because the first-order Ricci scalar perturbations can be treated as, in a sense, the classical source terms, similar to the approach used in GR. These ideas are implied by Li et al. in \cite{Li:2022pcy}. 

The partially geometric nature of Teukolsky's equation makes it particularly suitable for models beyond GR that do not derive from a specific Lagrangian, such as phenomenlogical models of quantum-gravity corrected BHs. For example, in LQG, dynamics of a BH is typically difficult to obtain from an effective Lagrangian. Indeed, the effective Lagrangian of LQG is remained unknown in the community. They often take Hamiltonian analysis for strictly spherically symmetric solutions, which limits the derivation of equations describing gravitational perturbations \cite{Lewandowski:2022zce,Gambini:2008dy,Kelly:2020uwj}.
In LQG, the researches of BHs are usually based on the Ashtekar variables \cite{ashtekar1986new}. 
In this scenario, the spacetime is reduced to one with spherical symmetry and rewritten in terms of Ashtekar variables. 
The quantum gravity is taken account basing on the Hamiltonian analysis, and the effective Lagrangian is absent \cite{husain2022fate}.  And solutions which are not directly derived from the Lagrangian also originates from studies of regular BHs \cite{Frolov:2016pav,Xiang:2013sza}.
Strictly speaking, perturbation equations for these 
solutions, even for minimally coupled massless scalar fields, are subtle. It is challenging to clarify how other fields interact with gravity without clear dynamics. However, within the framework of the Teukolsky's equation, handling gravitational perturbations becomes somewhat simpler. On one hand, these models are often more adept at estimating deviations in geometric quantities relative to GR, allowing for more accurate source terms. On the other hand, this approach can ultimately be reduced to calculations analogous to gravitational waves in EMRI systems, which have been extensively validated in gravitational wave studies.

Thus, for both BHs in modified gravity and those with quantum corrections, curvature perturbation studies offer an effective and unified approach for exploring perturbation properties of BHs beyond GR. 

\subsection{Modified Teukolsky's equation with $tr$ symmetry}
We consider four-dimensional asymptotically flat spherically-symmetric static metrics, which general form in Boyer–Lindquist coordinates is
\begin{equation}
\mathrm{d} s^2=G^2(r) \mathrm{d} t^2-\frac{1}{F^2(r)} \mathrm{d} r^2-r^2 \left(\mathrm{d} \theta^2+\operatorname{sin}^2 \theta\mathrm{d}\varphi^2\right),
\label{metric}
\end{equation}
with asymptotic conditions
\begin{equation}
F(r) \underset{r \rightarrow+\infty}{\longrightarrow} 1, \quad G(r) \underset{r \rightarrow+\infty}{\longrightarrow} 1.
\end{equation}
If we further have 
\begin{equation}
G^2(r)=F^2(r) \equiv \frac{\tilde{\Delta}(r)}{r^2},
\end{equation}
the spacetime is so-called $tr$-symmetric, which covers a significant number of models. Here we use tilde to distinguish the metric component functions from NP
directional derivative. 

We can choose a Kinnersley null tetrad
\begin{equation}
    l^{\mu}=\left( \frac{r^2}{\tilde{\Delta}(r)},1,0,0\right),n^{\mu}=\frac{1}{2}\left(1,-\frac{\tilde{\Delta}(r)}{r^2},0,0\right),m^{\mu}=\left(0,0, \frac{1}{\sqrt{2} r}, \frac{i \operatorname{csc}\theta}{\sqrt{2} r}\right),\bar{m}^{\mu}=\left(0,0, \frac{1}{\sqrt{2} r}, -\frac{i \operatorname{csc}\theta}{\sqrt{2} r}\right).
\end{equation}

In this tetrad, non-zero NP scalars are
\begin{equation}
\begin{aligned}
    \rho=-\frac{1}{r}, \mu=-\frac{\tilde\Delta }{2 r^3}, \gamma=\frac{-2\tilde{\Delta}+r\tilde{\Delta}'}{4r^3},\alpha=-\beta=-&\frac{\operatorname{cot}\theta}{2\sqrt{2}r},\Psi_{2}=-\frac{-12\tilde{\Delta}+6r\tilde{\Delta}'+2r^2-r^2\tilde{\Delta}'' }{12r^4},\\
    \Phi_{11}=-\frac{-4\tilde{\Delta}-4r\tilde{\Delta}'+2r^2+r^2\tilde{\Delta}}{8r^4}&,\Lambda=\frac{-2+\tilde{\Delta}''}{24r^2},
\end{aligned}
\end{equation}
where prime denotes radial derivative. Thus, this spacetime is  Petrov type-D, sharing the same condition in eq.(\ref{Sachs}). 

Similar to the GR case, one can take perturbation of eqs.(\ref{Teukolsky1},\ref{Teukolsky2},\ref{Teukolsky3}). The results are
\begin{equation}
\begin{aligned}
&\left(\bar{\delta}-4 \alpha+\pi\right)  \Psi_0{ }^{(1)}-(D-4 \rho-2 \epsilon)  \Psi_1{ }^{(1)}-3 \kappa^{(1)} \Psi_2{ } +2\kappa{}^{(1)}\Phi_{11}=S_1,  \\
&(\Delta-4 \gamma+\mu)  \Psi_0{ }^{(1)}-(\delta-4 \tau-2 \beta)  \Psi_1{ }^{(1)}-3 \sigma^{(1)} \Psi_2{ }-2\sigma{}^{(1)}\Phi_{11} =S_2,\\
&\left(D-\rho-\bar{\rho}-3 \epsilon+\bar{\epsilon}\right)  \sigma^{(1)}-\left(\delta-\tau+\bar{\pi}-\bar{\alpha}-3 \beta\right)  \kappa^{(1)}-\Psi_0{ }^{(1)}=0.
\end{aligned}
\label{modifiedpsi0eq}
\end{equation}

Notice that $\Phi_{01}=0$ in this tetrad and it is the only Ricci scalar term in eqs.({\ref{comrel}},\ref{NPuse2},\ref{NPuse3},\ref{NPuse1}), thus eq.(\ref{elieq}) also holds in this general case. We can also apply $\left(\delta+\bar{\pi}-\bar{\alpha}-3 \beta-4 \tau\right)$ and $\left(D-3 \epsilon+\bar{\epsilon}-4 \rho-\bar{\rho}\right)$ to these modified equations:
\begin{equation}
\begin{aligned}
    &\left(\delta+\bar{\pi}-\bar{\alpha}-3 \beta-4 \tau\right)\left(\left(\bar{\delta}-4 \alpha+\pi\right)  \Psi_0{ }^{(1)}-(D-4 \rho-2 \epsilon)  \Psi_1{ }^{(1)}-3 \kappa^{(1)} \Psi_2{ } +2\kappa{}^{(1)}\Phi_{11}\right)\\
    -&\left(D-3 \epsilon+\bar{\epsilon}-4 \rho-\bar{\rho}\right)\left((\Delta-4 \gamma+\mu)  \Psi_0{ }^{(1)}-(\delta-4 \tau-2 \beta)  \Psi_1{ }^{(1)}-3 \sigma^{(1)} \Psi_2{ }-2\sigma{}^{(1)}\Phi_{11}\right).
\end{aligned}
\label{eq.a}
\end{equation}
We use eq.(\ref{elieq}) with $p=2,q=-4$ to eliminate $\Psi_1{}^{(1)}$ terms. So we can transform eq.(\ref{eq.a}) into
\begin{equation}
\begin{aligned}
        &\left(\delta+\bar{\pi}-\bar{\alpha}-3 \beta-4 \tau\right)\left(\left(\bar{\delta}-4 \alpha+\pi\right)  \Psi_0{ }^{(1)}-3 \kappa^{(1)} \Psi_2{ } +2\kappa{}^{(1)}\Phi_{11}\right)\\
    -&\left(D-3 \epsilon+\bar{\epsilon}-4 \rho-\bar{\rho}\right)\left((\Delta-4 \gamma+\mu)  \Psi_0{ }^{(1)}-3 \sigma^{(1)} \Psi_2{ }-2\sigma{}^{(1)}\Phi_{11}\right).
\end{aligned}
\end{equation}
Introduce the operator
\begin{equation}
H_{GR}=(D-3 \epsilon+\bar{\epsilon}-4 \rho-\bar{\rho})(\Delta-4 \gamma+\mu)-(\delta+\bar{\pi}-\bar{\alpha}-3 \beta-4 \tau)(\bar{\delta}+\pi-4 \alpha),
\end{equation}
eq.(\ref{eq.a}) can be expressed as
\begin{equation}
    -H_{GR} \Psi_{0}{}^{(1)}+\left(\delta+\bar{\pi}-\bar{\alpha}-3 \beta-4 \tau\right)\left(-3 \kappa^{(1)} \Psi_2{ } +2\kappa{}^{(1)}\Phi_{11}\right)-\left(D-3 \epsilon+\bar{\epsilon}-4 \rho-\bar{\rho}\right)\left(-3 \sigma^{(1)} \Psi_2-2 \sigma^{(1)} \Phi_{11}\right)=-4\varpi T_0,
    \label{eq.b}
\end{equation}
where the source term shares the same form as the result in GR, but requires expressing the source term in terms of Ricci scalars.
Now we can focus on the left-hand side of the equation. The computations involving background quantities are straightforward, serving mainly as an effective potential terms in the corrected equations. However,  the perturbative spin coefficients increases the difficulty. Fortunately, the abundant gauge degrees of freedom of NP formalism provide a shortcut for further simplification, which is extensively used by Chandrasekhar \cite{Chandrasekhar:1985kt}. Let us take a class II transformation 
\begin{equation}
    n^{\mu} \rightarrow n^{\mu}, m^{\mu} \rightarrow m^{\mu}+b n^{\mu}, \bar{m}^{\mu} \rightarrow \bar{m}^{\mu}+\bar{b} n^{\mu}, l^{\mu} \rightarrow l^{\mu}+\bar{b} m^{\mu}+b \bar{m}^{\mu}+b \bar{b} n^{\mu},
\end{equation}
as shown in Appendix.\ref{NP}. Here $b$ is a perturbative quantity such that this transformation can only affect the first-order perturbations of all NP scalars. It is straightforward to verify that, given the background Kinnersley tetrad, perturbative NP scalars that change under class II transformations are only:
\begin{equation}
    \kappa^{(1)},\pi^{(1)},\epsilon^{(1)},\rho^{(1)},\sigma^{(1)},\beta^{(1)},\tau^{(1)},\Phi_{01}{}^{(1)},\Phi_{10}{}^{(1)},\Psi_1{}^{(1)}.
\end{equation}
In the equations we are discussing, only $ \kappa^{(1)},\sigma^{(1)},\Phi_{01}{}^{(1)},\Phi_{10}{}^{(1)}$ will influence the final result. So we can set $\sigma^{(1)}=0$ as a gauge fixing condition. So eq.(\ref{eq.b}) becomes
\begin{equation}
    -H_{GR} \Psi_{0}{}^{(1)}+\left(\delta+\bar{\pi}-\bar{\alpha}-3 \beta-4 \tau\right)\left(-3 \kappa^{(1)} \Psi_2{ } +2\kappa{}^{(1)}\Phi_{11}\right)=-4\varpi T_0.
\end{equation}
Notice that $\delta$ contains only angular derivative, and both $\Psi_2$ and $\Phi_{11}$ are angular independent scalars, so we can directly simplify the second term of left hand-side of this equation:
\begin{equation}
    \left(\delta+\bar{\pi}-\bar{\alpha}-3 \beta-4 \tau\right)\left(-3 \kappa^{(1)} \Psi_2{ } +2\kappa{}^{(1)}\Phi_{11}\right)=\left(-3\Psi_2+2\Phi_{11}\right)\left(\delta+\bar{\pi}-\bar{\alpha}-3 \beta-4 \tau\right)\kappa^{(1)}.
\end{equation}
Consider the third equation in eq.(\ref{modifiedpsi0eq}). Because of our gauge fixing condition, 
\begin{equation}
    -\left(\delta-\tau+\bar{\pi}-\bar{\alpha}-3 \beta\right) \kappa^{(1)}-\Psi_0^{(1)}=0.
\end{equation}
Notice that $\tau=0$,  we have
\begin{equation}
    \left(\delta+\bar{\pi}-\bar{\alpha}-3 \beta-4 \tau\right)\kappa^{(1)}=\left(\delta-\tau+\bar{\pi}-\bar{\alpha}-3 \beta\right) \kappa^{(1)}
\end{equation}
 Finally, we find that with a simple gauge fixing condition, the equations in this metric are once again decoupled, just like the story in GR:
\begin{equation}
(-H_{G R}+3\Psi_2-2\Psi_{11}) \Psi_0{ }^{(1)}=-4 \varpi T_0,
\end{equation}
or more specifically,
\begin{equation}
\left[(D-3 \epsilon+\bar{\epsilon}-4 \rho-\bar{\rho})(\Delta-4 \gamma+\mu)-(\delta+\bar{\pi}-\bar{\alpha}-3 \beta-4 \tau)(\bar{\delta}+\pi-4 \alpha)-3 \Psi_2+2\Phi_{11}\right] \Psi_0{ }^{(1)}=4 \varpi T_0.
\label{ModifiedTeukPsi0}
\end{equation}
In the case of $\Psi_4{}^{(1)}$, we can take a prime operation of eq.(\ref{ModifiedTeukPsi0}). The rusult is
\begin{equation}
\left[(\Delta+3 \gamma-\bar{\gamma}+4 \mu+\bar{\mu})(D+4 \epsilon-\rho)-(\bar{\delta}-\bar{\tau}+\bar{\beta}+3 \alpha+4 \pi)(\delta-\tau+4 \beta)-3 \Psi_2+2\Phi_{11}\right] \Psi_4^{(1)}=4 \varpi T_4.
\end{equation}

We can reformulate these two equations into a unified master equation form:
\begin{equation}
\begin{aligned}
   \left[ \frac{r^4}{\tilde{\Delta}}\frac{\partial^2}{\partial t^2}-\frac{1}{\operatorname{sin}^2 \theta}\frac{\partial^2}{\partial \varphi^2}-\tilde{\Delta}^{-s}\frac{\partial}{\partial r} \left(\tilde{\Delta}^{s+1} \frac{\partial}{\partial r}\right)-\frac{1}{\operatorname{sin}\theta}\frac{\partial}{\partial\theta}\left(\operatorname{sin}\theta\frac{\partial}{\partial\theta}\right)-2si\frac{\cos\theta}{\sin^2\theta}\frac{\partial}{\partial \varphi}+sr\left(4-\frac{r\tilde{\Delta}'}{\Delta}\right)\frac{\partial}{\partial t}\right.\\
    \left.+\left(s^2\cot^2\theta-s+(s+2)+2\frac{r\tilde{\Delta}'-\Delta}{r^2}-\frac{s+4}{2}\tilde{\Delta}''\right)\right]\psi=4\varpi r^2 T
\end{aligned}
\end{equation}
where
\begin{equation}
    \begin{aligned}
    s&=+2: & \quad &\psi=\Psi_0{}^{(1)} & \quad &T=2T_0,\\
    s&=-2: & \quad & \psi=\rho^{-4}\Psi_0{}^{(1)}& \quad &T=2\rho^{-4}T_4.
    \end{aligned}
\end{equation}
This equation is separable. Suppose
\begin{equation}
\psi=\int_{\omega} \sum_{lm} e^{-i \omega t} e^{i m \varphi} S_s(\theta) R_s(r)d\omega,
\end{equation}
we can get the radial and angular equations:
\begin{equation}
\frac{1}{\sin \theta} \frac{d}{d \theta}\left(\sin \theta \frac{d S_s}{d \theta}\right)+ \left(-\frac{m^2}{\sin ^2 \theta}-\frac{2 m s \cos \theta}{\sin ^2 \theta}-s^2 \cot ^2 \theta+s+\Lambda\right) S_s=0.
\end{equation}
\begin{equation}
\tilde{\Delta}^{-s} \frac{d}{d r}\left(\tilde{\Delta}^{s+1} \frac{d R_s}{d r}\right)+\left(\frac{\omega^2 r^4- i s\tilde{\Delta}' \omega r^2 }{\tilde{\Delta}}+4 i s \omega r-\Lambda-\left((s+2)+2\frac{r\tilde{\Delta}'-\tilde{\Delta}}{r^2}-\frac{s+4}{2}\tilde{\Delta}''\right)\right) R_s=T_{slm\omega}.
\label{RadialMod}
\end{equation}
Here $S_s(\theta)$ is the spin-weighted spherical harmonic, and
\begin{equation}
A=(l-s)(l+s+1)
\end{equation}
is its eigenvalue.

For comparasion, in the Schwarzschild case ($\tilde{\Delta}(r)=r^2-2 M r$), the Teukolsky master equation and its corresponding radial ordinary differential equation in Boyer-Lindquist coordinates are given by \cite{Teukolsky:1973ha}:

\begin{equation} 
\begin{aligned} {\frac{r^4}{\tilde{\Delta}}\ \frac{\partial^2 \psi}{\partial t^2}-\frac{1}{\sin ^2 \theta} \frac{\partial^2 \psi}{\partial \varphi^2}}-\tilde{\Delta}^{-s} \frac{\partial}{\partial r}\left(\tilde{\Delta}^{s+1} \frac{\partial \psi}{\partial r}\right)-\frac{1}{\sin \theta} \frac{\partial}{\partial \theta}\left(\sin \theta \frac{\partial \psi}{\partial \theta}\right)-2 si\frac{ \cos \theta}{\sin ^2 \theta}\frac{\partial \psi}{\partial \varphi} \\ \quad-2 s\left(\frac{Mr^2}{\tilde{\Delta}}-r\right) \frac{\partial \psi}{\partial t}+\left(s^2 \cot ^2 \theta-s\right) \psi=4 \pi r^2 T, 
\end{aligned} 
\end{equation}

\begin{equation} \tilde{\Delta}^{-s} \frac{d}{d r}\left(\tilde{\Delta}^{s+1} \frac{d R_s}{d r}\right)+\left(\frac{\omega^2 r^4-2 i s(r-M) \omega r^2}{\tilde{\Delta}}+4 i s \omega r-\Lambda\right) R_s=0.
\label{RadialGR}
\end{equation}

It is straightforward to verify that all our results are consistent with those obtained in GR. We find that in the $tr$-symmetric spacetime, the curvature perturbations still form a set of decoupled equations. Compared to the GR case from eqs.(\ref{RadialGR},\ref{RadialMod}), these equations include not only terms equivalent to classical source terms but also corrections to the effective potential,
\begin{equation}
    V_{\text{eff}}=(s+2)+2\frac{r\tilde{\Delta}'-\tilde{\Delta}}{r^2}-\frac{s+4}{2}\tilde{\Delta}'',
\end{equation} 
which comes from the non-Ricci-flat nature of the background. Once the equations are separated into variables as in GR, this new effective potential term will appear even in homogeneous cases, affecting the propagation of gravitational perturbations in this background.

\subsection{More general cases}
Once the $tr$ symmetry vanishes, that is, metric takes the form of eq.(\ref{metric}), there also exists a Kinnersley null tetrad
\begin{equation}
    l^{\mu}=\left( \frac{1}{G^2(r)},\frac{F(r)}{G(r)},0,0\right),n^{\mu}=\frac{1}{2}\left(1,G(r)F(r),0,0\right),m^{\mu}=\left(0,0, \frac{1}{\sqrt{2} r}, \frac{i \operatorname{csc}\theta}{\sqrt{2} r}\right),\bar{m}^{\mu}=\left(0,0, \frac{1}{\sqrt{2} r}, -\frac{i \operatorname{csc}\theta}{\sqrt{2} r}\right).
\end{equation}
We notice that, compared to the previously discussed case, there are two additional non-vanishing NP scalars:
\begin{equation}
    \Phi_{00}=-\frac{F\left(-G F^{\prime}+F G^{\prime}\right)}{r G^3}, \quad \Phi_{22}=-\frac{F G\left(-G F^{\prime}+FG^{\prime}\right)}{4 r}.
\end{equation}
Therefore, when handling eqs.(\ref{Teukolsky1},\ref{Teukolsky2},\ref{Teukolsky3}), we get:
\begin{equation}
\begin{aligned}
&\left(\bar{\delta}-4 \alpha+\pi\right)  \Psi_0{ }^{(1)}-(D-4 \rho-2 \epsilon)  \Psi_1{ }^{(1)}-3 \kappa^{(1)} \Psi_2{ } +2\kappa{}^{(1)}\Phi_{11}-\left(\delta+\bar{\pi}-2\bar{\alpha}-2\beta\right)^{(1)}\Phi_{00}=S_1,  \\
&(\Delta-4 \gamma+\mu)  \Psi_0{ }^{(1)}-(\delta-4 \tau-2 \beta)  \Psi_1{ }^{(1)}-3 \sigma^{(1)} \Psi_2{ }-2\sigma{}^{(1)}\Phi_{11}+\bar{\lambda}^{(1)}\Phi_{00} =S_2,\\
&\left(D-\rho-\bar{\rho}-3 \epsilon+\bar{\epsilon}\right)  \sigma^{(1)}-\left(\delta-\tau+\bar{\pi}-\bar{\alpha}-3 \beta\right)  \kappa^{(1)}-\Psi_0{ }^{(1)}=0.
\end{aligned}
\end{equation}
The presence of $\delta^{(1)}$ here renders the gauge-fixing technique used previously nearly inapplicable. Indeed, the perturbation of directional derivatives can only be fixed according to \cite{Loutrel:2020wbw}:

\begin{equation}
\begin{aligned}
D^{(1)} & =-\frac{1}{2} h_{l l} \Delta^{(0)}, \\
\Delta^{(1)} & =-\frac{1}{2} h_{n n} D^{(0)}-h_{l n} \Delta^{(0)}, \\
\delta^{(1)} & =-h_{n m} D^{(0)}-h_{l m} \Delta^{(0)} +\frac{1}{2} h_{m \bar{m}} \delta^{(0)}+\frac{1}{2} h_{m m} \bar{\delta}^{(0)}.
\end{aligned}
\end{equation}
To date, to the best of the author's knowledge, there has not been an effective gauge fixing schemes, such as radiation gauge or Bondi gauge in the study of metric reconstruction problems \cite{Loutrel:2020wbw,Ori:2002uv}, for metric perturbations that can adequately addresses this issue. So, in principle, it is difficult to obtain decoupled differential equations for the Weyl scalars in such spherically symmetric spacetimes. 

Nevertheless, the methods we previously derived are not without significance in solving such problems. Current tests of GR suggest that the $tr$ symmetry might be slightly violated by new physics. In this case, the $\Phi_{00}$ term could be treated as source term on a $tr$ symmetric background, just like the treatment in \cite{Li:2022pcy}. 

\section{Comparison with other methods} \label{Sec3}
There are other workarounds to the curvature perturbation problems beyond GR in a unified manner. In \cite{Arbey:2021jif,Arbey:2021yke}, Arbey et al. derived the Teukolsky's equations for massless fields of all spins in general spherically-symmetric and
static metrics. The form of the equations they obtain is entirely identical to the Type D vacuum cases in GR. This approach assumes that the coupling of massless fields to the background is identical to that in GR, implying that the equations in the NP formalism should be the same.

This assumption holds for non-gravitational fields, but it is incorrect to regard the gravitational field equations as those in the Type D vacuum cases. The Ricci scalars of background cannot be ignored.  Indeed, even in the perturbation theory of spherically symmetric BHs in the Einstein-Maxwell theory, the background curvature affects the equations of motion \cite{Chandrasekhar:1985kt,Chandrasekhar:1979iz,Giorgi:2019kjt,Giorgi:2018ldn}. For example, S.Chandrasekhar modified $\kappa^{(1)},\sigma^{(1)}$ terms in eq.(\ref{psi0eq}) on Reissner–Nordstr\"{o}m background by non-zero Ricci scalars, which is similar to our calculations in eq.(\ref{eq.a}). He simply expressed the Ricci scalars as contributions from specific energy-momentum tensors, i.e., $2\phi_1\bar{\phi}_1$, where $\phi_1=\frac{1}{2}F_{\mu\nu}\left(l^{\mu}n^{\nu}+\bar{m}^{\mu}m^{\nu}\right)$ is related to the antisymmetric Maxwell tensor $F_{\mu\nu}$. This implies that, at least for gravitational perturbations, the Ricci-ignored method in \cite{Arbey:2021jif} actually conflicts with the existing perturbation theory for Reissner–Nordstr\"{o}m BHs. Therefore, although this method can yield decoupled curvature perturbation equations in any static spherically symmetric spacetime, its application does carry certain risks.

Another formalism is developed by Li et al. in \cite{Li:2022pcy}. It is based on two-parameter
expansion for modified gravity theories. The first one,  $\zeta$, is a dimensionless parameter that characterizes the strength of the correction to GR, and
the second one, $\epsilon$, is a dimensionless parameter that describes the size of the gravitational perturbations. Since this formalism extends the calculation on non-Ricci-flat, Petrov type D BH backgrounds to the non-Ricci-flat, Petrov type I BH backgrounds, it has been used in the study of BH perturbation theories beyond GR, such as higher-derivative gravity and dynamical Chern-Simons gravity for recent years \cite{Cano:2023tmv,Wagle:2023fwl,Richards:2023xsr}.

Our method is inspired by this approach, which allows for the careful introduction of suitable gauge fixing during calculations, originating from Chandrasekhar's treatment of BH perturbation theory. However, in the study of static spherically symmetric cases, our method demonstrates reasonable analytical properties compared to the two-parameter expansion method. Following \cite{Li:2022pcy}, for non-GR
non-Ricci-flat Petrov type D spacetimes (this is also relevant to the discussions in this paper), we can introduce 
\begin{equation}
\begin{aligned}
& \mathcal{E}_1=\delta-\tau+\bar{\pi}-\bar{\alpha}-3 \beta-\frac{1}{\Psi_2} \delta \Psi_2, \\
& \mathcal{E}_2=D-\rho-\bar{\rho}-3 \varepsilon+\bar{\varepsilon}-\frac{1}{\Psi_2} D \Psi_2, \\
& \mathcal{E}_3=\bar{\delta}+3 \alpha+\bar{\beta}+\pi-\bar{\tau}-\frac{1}{\Psi_2} \bar{\delta} \Psi_2,\\
& \mathcal{E}_4=\Delta+\mu+\bar{\mu}+3 \gamma-\bar{\gamma}-\frac{1}{\Psi_2} \Delta \Psi_2,
\end{aligned}
\label{Ei}
\end{equation}
and
\begin{equation}
\begin{aligned}
&\begin{aligned}
& F_1 \equiv \bar{\delta}-4 \alpha+\pi, \quad F_2 \equiv \Delta-4 \gamma+\mu, \\
& J_1 \equiv D-2 \varepsilon-4 \rho, \quad J_2 \equiv \delta-4 \tau-2 \beta, \\
& E_1 \equiv \delta-\tau+\bar{\pi}-\bar{\alpha}-3 \beta, \\
& E_2 \equiv D-\rho-\bar{\rho}-3 \varepsilon+\bar{\varepsilon},
\end{aligned}\\
&\begin{aligned}
& F_3 \equiv \delta+4 \beta-\tau, \quad F_4 \equiv D+4 \varepsilon-\rho ,\\
& J_3 \equiv \Delta+2 \gamma+4 \mu, \quad J_4 \equiv \bar{\delta}+4 \pi+2 \alpha ,\\
& E_3 \equiv \bar{\delta}+3 \alpha+\bar{\beta}+\pi-\bar{\tau}, \\
& E_4 \equiv \Delta+\mu+\bar{\mu}+3 \gamma-\bar{\gamma}.
\end{aligned}
\end{aligned}
\end{equation}
The decoupled equations are
\begin{equation}
\begin{aligned}
&\left(\mathcal{E}_2 F_2-\mathcal{E}_1 F_1-3 \Psi_2\right)\Psi_0{}^{(1)}=\mathcal{E}_2 S_2-\mathcal{E}_1 S_1,\\
&\left(\mathcal{E}_4 F_4-\mathcal{E}_3 F_3-3 \Psi_2\right)\Psi_4{}^{(1)}=\mathcal{E}_4 S_4-\mathcal{E}_3 S_3.
\end{aligned}
\label{Chen}
\end{equation}

The main issue with this set of equations arises from operators $\mathcal{E}_i$. As we can see from eq.(\ref{Ei}), all these operators contain a term with a denominator of $\Psi_2$. Therefore, as long as the directional derivative of $\Psi_2$ cannot be expressed as a product involving other NP scalars, just like eq.(\ref{Dpsi22}), there exists a singularity in the differential equations arising from this term $\Psi_2=0$. In our cases, it is
\begin{equation}
    -12\tilde{\Delta}+6r\tilde{\Delta}'+2r^2-r^2\tilde{\Delta}''=0.
\end{equation}

Obviously, it cannot produce the the same singularities as $\tilde{\Delta}=0$. On the one hand, $\Psi_2=0$ does not correspond to any special physical phenomena, such as the existence of physical singularities or horizons. On the other hand, analysis of the background metric and its curvature components shows that it is only singularities from $\tilde{\Delta}=0$ that should appear due to divergences in geometric quantities in Boyer–Lindquist coordinates. In fact, Chandrasekhar's analysis of Reissner–Nordstr\"{o}m BHs \cite{Chandrasekhar:1985kt,Chandrasekhar:1979iz} indicates that all singularities in the equations he derived align with our discussion, while eq.(\ref{Chen}) cannot reproduce this property. This implies that the formalism in \cite{Li:2022pcy} may encounter ``fake poles'' in practical applications. Given that the poles of the equations are essential to BH perturbation theory—both in analytical and numerical contexts \cite{Sasaki:2003xr,Berti:2009kk,Chia:2020yla}—this subtle issue is likely to induce unnecessary complications. 

In contrast, our equations do not manifest these problems. Therefore, while the domain of strict validity for our equations is limited, these equations offer distinct advantages over existing formalisms in the relevant contexts.

\section{SUMMARY AND DISCUSSION}\label{Sec4}
Inspired by recent developments of modified Teukolsky formalism \cite{Li:2022pcy}, we develop a set of new decopuled equations of $\Psi_0$ and $\Psi_4$ which are suitable for generic spherically-symmetric static metrics, especially for spacetimes with $tr$ symmetry. We use the similar gauge fixing method in \cite{Chandrasekhar:1985kt,Li:2022pcy}, taking a more careful approach in addressing the changes in NP scalars induced by gauge transformations, thereby avoiding the occurrence of fake poles. 

When applying our formalism, the treatment of the source term will effectively reflect the dynamical deviations from GR. From a certain perspective, it is desirable for these source terms to be established through alternative methods prior to analysis of our formalism, rather than relying on techniques such as metric reconstruction in \cite{Li:2022pcy,Wagle:2023fwl}. Indeed, even in GR, metric reconstruction presents considerable complexity, regardless the challenges posed by other modified gravity theories \cite{Spiers:2023cip,Hollands:2024iqp,Green:2019nam}. A compromise solution, as discussed in \cite{Wagle:2023fwl}, involves further expansion of the background metric, treating it as both GR background and non-GR perturbative components, and subsequently obtains appropriate source terms in the GR background. Thus, our formalism requires further investigation and validation within the context of specific modified gravity theories. We leave this work for future study.

However, this work is particularly beneficial for studies aimed at understanding GWs from BHs with quantum corrections. According to our formalism, the dynamics could be regarded as a EMRI system with some given fields on the non-Ricci-flat background. These source terms should physically correspond to quantum fluctuations in spacetime, and reasonable estimates can be obtained based on certain arguments, just like what happens in loop quantum cosmology. We hope that our work will assist more researchers in this area, and we will discuss a specific case in the future.

\section {Acknowledgments}  
 This work is supported by the National Key Research and Development Program of China Grant No.2020YFC2201502, grants from NSFC (grant No. 12250010, 12475065), Key Research Program of Frontier Sciences, CAS, Grant NO. ZDBS-LY-7009. 

\begin{appendices}
\section{NP Formalism (Full Version)}\label{NP}
\setcounter{equation}{0}
\renewcommand{\theequation}{A.\arabic{equation}}
In Section.\ref{review}, we have introduced the orthogonality relations for the tetrad basis vectors in NP formalism. We denote these vectors as
\begin{equation}
e_a^\mu=\left(l^\mu, n^\mu, m^\mu, \bar{m}^\mu\right) \quad(a=1,2,3,4).
\end{equation}
Thus the orthogonal condition can be further shown as $g_{\mu \nu}=e_\mu^a e_\nu^b \eta_{a b}$, where
\begin{equation}
\eta_{a b}=\eta^{a b}=\left(\begin{array}{cccc}
0 & 1 & 0 & 0 \\
1 & 0 & 0 & 0 \\
0 & 0 & 0 & -1 \\
0 & 0 & -1 & 0
\end{array}\right).
\end{equation}

Spin coefficients, also called Ricci rotation-coefficients, which could be regarded as the Christoffel symbols in tetrad formalism, are defined as 
\begin{equation}
\gamma_{cab}\equiv e_{c}^{\kappa} e_{b}^{\mu}\nabla_{\mu}e_{a \kappa},
\end{equation}
where 
\begin{equation}
\eta_{ab} e_{\mu}^{a}=e_{b \mu}, \quad \eta^{ab} e_{a \mu}=e^{b}_{\mu}.
\end{equation}

There are 12 complex spin coefficients in NP formalism, which are defined as 
\begin{equation}
\begin{aligned}
&\kappa=\gamma_{311} ; &\quad& \rho=\gamma_{314} ; &\quad& \varepsilon=\frac{1}{2}\left(\gamma_{211}+\gamma_{341}\right); \\
&\sigma=\gamma_{313} ; &\quad& \mu=\gamma_{243} ; &\quad& \gamma=\frac{1}{2}\left(\gamma_{212}+\gamma_{342}\right) ;\\
&\lambda=\gamma_{244} ; &\quad& \tau=\gamma_{312} ; &\quad& \alpha=\frac{1}{2}\left(\gamma_{214}+\gamma_{344}\right) ; \\
&\nu=\gamma_{242} ; &\quad& \pi=\gamma_{241} ; &\quad& \beta=\frac{1}{2}\left(\gamma_{213}+\gamma_{343}\right) .
\end{aligned}
\end{equation}

The first set of equations that can be obtained is the directional derivatives for NP tetrad:
\begin{equation}
\begin{aligned}
D l^\mu  =(\epsilon+\bar{\epsilon}) l^a-\bar{\kappa} m^\mu-\kappa \bar{m}^\mu ,&\quad&\Delta l^\mu  =(\gamma+\bar{\gamma}) l^\mu-\bar{\tau} m^\mu-\tau \bar{m}^\mu, \\
\delta l^\mu  =(\beta+\bar{\alpha}) l^\mu-\bar{\rho} m^\mu-\sigma \bar{m}^\mu, &\quad&\bar{\delta} l^\mu  =(\alpha+\bar{\beta}) l^\mu-\bar{\sigma} m^\mu-\rho \bar{m}^\mu, \\
D n^\mu  =-(\epsilon+\bar{\epsilon}) n^\mu+\pi m^\mu+\bar{\pi} \bar{m}^\mu,&\quad&\Delta n^\mu  =-(\gamma+\bar{\gamma}) n^\mu+v \bar{m}^\mu+\bar{v} \bar{m}^\mu, \\
\delta n^\mu  =-(\beta+\bar{\alpha}) n^\mu+\mu m^\mu+\bar{\lambda} \bar{m}^\mu, &\quad&\bar{\delta} n^\mu  =-(\alpha+\bar{\beta}) n^\mu+\lambda m^\mu+\bar{\mu} \bar{m}^\mu, \\
D m^\mu  =(\epsilon-\bar{\epsilon}) m^\mu+\bar{\pi} l^\mu-\kappa n^\mu,&\quad&\Delta m^\mu  =(\gamma-\bar{\gamma}) m^\mu+\bar{v} l^\mu-\tau n^\mu, \\
\delta m^\mu  =(\beta-\bar{\alpha}) m^\mu+\bar{\lambda} l^\mu-\sigma n^\mu, &\quad&\bar{\delta} m^\mu  =(\alpha-\bar{\beta}) m^\mu+\bar{\mu} l^\mu-\rho n^\mu, \\
D \bar{m}^\mu  =(\bar{\epsilon}-\epsilon) \bar{m}^\mu+\pi l^\mu-\bar{\kappa} n^\mu,&\quad&\Delta \bar{m}^\mu  =(\bar{\gamma}-\gamma) \bar{m}^\mu+v l^\mu-\bar{\tau} n^\mu, \\
\delta \bar{m}^\mu =(\bar{\alpha}-\beta) \bar{m}^\mu+\mu l^\mu-\bar{\rho} n^\mu,&\quad&\bar{\delta} \bar{m}^\mu  =(\bar{\beta}-\alpha) \bar{m}^\mu+\lambda l^\mu-\bar{\sigma} n^\mu.
\end{aligned}
\end{equation}

Besides, one can further get the relations between commutation relations and spin coefficients :
\begin{equation}
\begin{aligned}
(\Delta D-D \Delta) \phi & =[(\gamma+\bar{\gamma}) D+(\epsilon+\bar{\epsilon}) \Delta-(\bar{\tau}+\pi) \delta-(\tau+\bar{\pi}) \bar{\delta}] \phi, \\
(\delta D-D \delta) \phi & =[(\bar{\alpha}+\beta-\bar{\pi}) D+\kappa \Delta-(\bar{\rho}+\epsilon-\bar{\epsilon}) \delta-\sigma \delta ] \phi,
\\
(\delta \Delta-\Delta \delta) \phi & =[-\bar{v} D-(\bar{\alpha}+\beta-\tau) \Delta+(\mu-\gamma+\bar{\gamma}) \delta+\bar{\lambda}\bar{ \delta}] \phi, \\
(\bar{\delta} \delta-\delta \bar{\delta}) \phi & =[(\bar{\mu}-\mu) D+(\bar{\rho}-\rho) \Delta-(\bar{\beta}-\alpha) \delta-(\bar{\alpha}-\beta) \bar{\delta}] \phi.
\end{aligned}
\label{comrel}
\end{equation}

In NP formalism, the relationship among the Riemann tensor, Weyl tensor, and Ricci tensor remains unchanged in tetrad notation:

\begin{equation}
R_{a b c d}=C_{a b c d}  -\frac{1}{2}\left(\eta_{a c} R_{b d}-\eta_{b c} R_{a d}-\eta_{a d} R_{b c}+\eta_{b d} R_{a c}\right)  +\frac{1}{6}\left(\eta_{a c} \eta_{b d}-\eta_{a d} \eta_{b c}\right) R .
\end{equation}
So Curvature tensor can be expressed by 5 complex Weyl scalars
\begin{equation}
\begin{aligned}
& \Psi_0=-C_{1313}=-C_{\alpha \beta \gamma \delta} l^\alpha m^\beta l^\gamma m^\delta, \\
& \Psi_1=-C_{1213}=-C_{\alpha \beta \gamma \delta} l^\alpha n^\beta l^\gamma m^\delta, \\
& \Psi_2=-C_{1342}=-C_{\alpha \beta \gamma \delta} l^\alpha m^\beta \bar{m}^\gamma n^\delta, \\
& \Psi_3=-C_{1242}=-C_{\alpha \beta \gamma \delta} l^\alpha n^\beta \bar{m}^\gamma n^\delta, \\
& \Psi_4=-C_{2424}=-C_{\alpha \beta \gamma \delta} n^\alpha \bar{m}^\beta n^\gamma \bar{m}^\delta,
\end{aligned}
\end{equation}
and 10 NP Ricci scalars
\begin{equation}
\begin{aligned}
& \Phi_{00}=-\frac{1}{2} R_{11}=-\frac{1}{2} R_{\mu \nu} l^\mu l^\nu , \Phi_{01}=-\frac{1}{2} R_{13}=-\frac{1}{2} R_{\mu \nu} l^\mu m^\nu, \Phi_{10}=-\frac{1}{2} R_{14}=-\frac{1}{2} R_{\mu \nu} l^\mu \bar{m}^\nu, \\
& \Phi_{11}=-\frac{1}{4}\left(R_{12}+R_{34}\right)=-\frac{1}{2} R_{\mu \nu}\left(l^\mu n^\nu+m^\mu \bar{m}^\nu\right), \\
& \Phi_{02}=-\frac{1}{2} R_{33}=-\frac{1}{2} R_{\mu \nu} m^\mu m^\nu, \Phi_{12}=-\frac{1}{2} R_{23}=-\frac{1}{2} R_{\mu \nu} n^\mu m^\nu, \\
& \Phi_{20}=-\frac{1}{2} R_{44}=-\frac{1}{2} R_{\mu \nu} \bar{m}^\mu \bar{m}^\nu, \Phi_{21}=-\frac{1}{2} R_{24}=-\frac{1}{2} R_{\mu \nu} n^\mu \bar{m}^\nu, \\
& \Phi_{22}=-\frac{1}{2} R_{22}=-\frac{1}{2} R_{\mu \nu} n^\mu n^\nu, \Lambda=R / 24.
\end{aligned}
\end{equation}
Using these scalars, one can get Ricci identities according to 
\begin{equation}
R_{a b c d}= -\gamma_{a b c, d}+\gamma_{a b d, c}+\gamma_{a b f}\left(\gamma^f{ }_{c d}-\gamma^f{ }_{d c}\right)  +\gamma^f{ }_{a c} \gamma_{b f d}-\gamma^f{ }_{a d} \gamma_{b f c},
\end{equation}
where $\gamma_{a b c, d} \equiv \gamma_{a b c, \mu} e_d^\mu$. The explicit form is
\setcounter{equation}{0}
\renewcommand{\theequation}{A.12.\arabic{equation}} 

\begin{align}
& D \rho-\bar{\delta} \kappa=\left(\rho^2+\sigma \bar{\sigma}\right)+(\epsilon+\bar{\epsilon}) \rho-\bar{\kappa} \tau-\kappa(3 \alpha+\bar{\beta}-\pi)+\Phi_{00}, \\
& D \alpha-\bar{\delta} \epsilon=(\rho+\bar{\epsilon}-2 \epsilon) \alpha+\beta \bar{\sigma}-\bar{\beta} \epsilon-\kappa \lambda-\bar{\kappa} \gamma+(\epsilon+\rho) \pi+\Phi_{10}, \\
& D \lambda-\bar{\delta} \pi=(\rho-3 \epsilon+\bar{\epsilon}) \lambda+\bar{\sigma} \mu+(\pi+\alpha-\bar{\beta}) \pi-\nu \bar{\kappa}+\Phi_{20}, \\
& \Delta \mu-\delta v=-(\mu+\gamma+\bar{\gamma}) \mu-\lambda \bar{\lambda}+\bar{v} \pi+(\bar{\alpha}+3 \beta-\tau) \nu-\Phi_{22}, \\
& \Delta \beta-\delta \gamma=(\bar{\alpha}+\beta-\tau) \gamma-\mu \tau+\sigma \nu+\epsilon \bar{\nu}+(\gamma-\bar{\gamma}-\mu) \beta-\alpha \bar{\lambda}-\Phi_{12}, \\
& \Delta \sigma-\delta \tau=-(\mu-3 \gamma+\bar{\gamma}) \sigma-\bar{\lambda} \rho-(\tau+\beta-\bar{\alpha}) \tau+\kappa \bar{\nu}-\Phi_{02}, \\
& D \sigma-\delta \kappa=(\rho+\bar{\rho}+3 \epsilon-\bar{\epsilon}) \sigma-(\tau-\bar{\pi}+\bar{\alpha}+3 \beta) \kappa+\Psi_0, \\
& D \beta-\delta \epsilon=(\alpha+\pi) \sigma+(\bar{\rho}-\bar{\epsilon}) \beta-(\mu+\gamma) \kappa-(\bar{\alpha}-\bar{\pi}) \epsilon+\Psi_1 \label{NPuse2}, \\
& \Delta \alpha-\bar{\delta} \gamma=(\rho+\epsilon) v-(\tau+\beta) \lambda+(\bar{\gamma}-\bar{\mu}) \alpha+(\bar{\beta}-\bar{\tau}) \gamma-\Psi_3 ,\\
& \Delta \lambda-\bar{\delta} v=-(\mu+\bar{\mu}+3 \gamma-\bar{\gamma}) \lambda+(3 \alpha+\bar{\beta}+\pi-\bar{\tau}) \nu-\Psi_4, \\
& D \tau-\Delta \kappa=(\tau+\bar{\pi}) \rho+(\bar{\tau}+\pi) \sigma+(\epsilon-\bar{\epsilon}) \tau-(3 \gamma+\bar{\gamma}) \kappa+\Psi_1+\Phi_{01}, \label{NPuse1} \\
& \delta \lambda-\bar{\delta} \mu=(\rho-\bar{\rho}) \nu+(\mu-\bar{\mu}) \pi+(\alpha+\bar{\beta}) \mu+\bar{\alpha}-3 \beta) \lambda-\Psi_3+\Phi_{21}, \\
& D \nu-\Delta \pi=(\pi+\bar{\tau}) \mu+(\bar{\pi}+\tau) \lambda+(\gamma-\bar{\gamma}) \pi-(3 \epsilon+\bar{\epsilon}) \nu+\Psi_3+\Phi_{21}, \\
& \delta \rho-\bar{\delta} \sigma=(\bar{\alpha}+\beta) \rho-(3 \alpha-\bar{\beta}) \sigma+(\rho-\bar{\rho}) \tau+(\mu-\bar{\mu}) \kappa-\Psi_1+\Phi_{01}, \label{NPuse3}\\
& D \gamma-\Delta \epsilon=(\tau+\bar{\pi}) \alpha+(\bar{\tau}+\pi) \beta-(\epsilon+\bar{\epsilon}) \gamma-(\gamma+\bar{\gamma}) \epsilon+\tau \pi-\nu \kappa  +\Psi_2-\Lambda+\Phi_{11}, \\
& \delta \alpha-\bar{\delta} \beta=\mu \rho-\lambda \sigma+\alpha \bar{\alpha}+\beta \bar{\beta}-2 \alpha \beta+(\rho-\bar{\rho}) \gamma+(\mu-\bar{\mu}) \epsilon -\Psi_2+\Lambda+\Phi_{11}, \\
& D \mu-\delta \pi=(\bar{\rho}-\epsilon-\bar{\epsilon}) \mu+\sigma \lambda+(\bar{\pi}-\bar{\alpha}+\beta) \pi-\nu \kappa+\Psi_2+2 \Lambda, \\
& \Delta \rho-\bar{\delta} \tau=(\gamma+\bar{\gamma}-\bar{\mu}) \rho-\sigma \lambda+(\bar{\beta}-\alpha-\bar{\tau}) \tau+\nu \kappa-\Psi_2-2 \Lambda.
\end{align}

Bianchi identities $R_{\alpha \beta[\gamma \delta ; \mu]}=0$ in NP formalism take the form:
\setcounter{equation}{0}
\renewcommand{\theequation}{A.13.\arabic{equation}} 

\begin{align}
D \Psi_1-\bar{\delta} \Psi_0-D \Phi_{01}+  \delta \Phi_{00} & =(\pi-4 \alpha) \Psi_0+2(2 \rho+\epsilon) \Psi_1-3 \kappa \Psi_2\notag\\
& -(\bar{\pi}-2 \bar{\alpha}-2 \beta) \Phi_{00}-2(\bar{\rho}+\epsilon) \Phi_{01}  -2 \sigma \Phi_{10}+2 \kappa \Phi_{11}+\bar{\kappa} \Phi_{02}, \\
\Delta \Psi_0-\delta \Psi_1+D \Phi_{02}-  \delta \Phi_{01}  &=(4 \gamma-\mu) \Psi_0-2(2 \tau+\beta) \Psi_1+3 \sigma \Psi_2 \notag\\
& -\bar{\lambda} \Phi_{00}+2(\bar{\pi}-\beta) \Phi_{01}+2 \sigma \Phi_{11} +(\bar{\rho}+2 \epsilon-2 \bar{\epsilon}) \Phi_{02}-2 \kappa \Phi_{12},\\
D \Psi_2-\bar{\delta} \Psi_1+\Delta \Phi_{00}-  \bar{\delta}\Phi_{01}+2 D \Lambda \notag& =-\lambda \Psi_0+2(\pi-\alpha) \Psi_1+3 \rho \Psi_2-2 \kappa \Psi_3\notag\\
& +(2 \gamma+2 \bar{\gamma}-\bar{\mu}) \Phi_{00}-2(\alpha+\bar{\tau}) \Phi_{01} -2 \tau \Phi_{10}+2 \rho \Phi_{11}+\bar{\sigma} \Phi_{02},\\
\Delta \Psi_1-\delta \Psi_2-\Delta \Phi_{01}+\bar{\delta} \Phi_{02}-  2 \delta \Lambda & =\nu \Psi_0+2(\gamma-\mu) \Psi_1-3 \tau \Psi_2+2 \sigma \Psi_3 \notag\\ 
& -\bar{\nu} \Phi_{00}+2(\bar{\mu}-\gamma) \Phi_{01} +(2 \alpha+\bar{\tau}-2 \bar{\beta}) \Phi_{02}+2 \tau \Phi_{11}-2 \rho \Phi_{12}, \label{Dpsi2}\\
D \Psi_3-\bar{\delta} \Psi_2-D \Phi_{21}+\delta \Phi_{20}-  2 \bar{\delta} \Lambda & =-2 \lambda \Psi_1+3 \pi \Psi_2+2(\rho-\epsilon) \Psi_3-\kappa \Psi_4\notag \\ 
& +2 \mu \Phi_{10}-2 \pi \Phi_{11}-(2 \beta+\bar{\pi}-2 \bar{\alpha}) \Phi_{20} -2(\bar{\rho}-\epsilon) \Phi_{21}+\bar{\kappa} \Phi_{22},
\label{deltapsi2}\\
\Delta \Psi_2-\delta \Psi_3+D \Phi_{22}-\delta \Phi_{21}+  2 \Delta \Lambda& =2 \nu \Psi_1-3 \mu \Psi_2+2(\beta-\tau) \Psi_3+\sigma \Psi_4 \notag\\
& -2 \mu \Phi_{11}-\bar{\lambda} \Phi_{20}+2 \pi \Phi_{12} +2(\beta+\bar{\pi}) \Phi_{21}+(\bar{\rho}-2 \epsilon-2 \bar{\epsilon}) \Phi_{22},\\
D \Psi_4-\bar{\delta} \Psi_3+\Delta \Phi_{20}-  \bar{\delta} \Phi_{21} & =-3 \lambda \Psi_2+2(\alpha+2 \pi) \Psi_3+(\rho-4 \epsilon) \Psi_4 \notag\\
& +2 \nu \Phi_{10}-2 \lambda \Phi_{11}-(2 \gamma-2 \bar{\gamma}+\bar{\mu}) \Phi_{20}  -2(\bar{\tau}-\alpha) \Phi_{21}+\bar{\sigma} \Phi_{22}, \\
\Delta \Psi_3-\delta \Psi_4-\Delta \Phi_{21}+  \bar{\delta} \Phi_{22} & =3 \nu \Psi_2-2(\gamma+2 \mu) \Psi_3+(4 \beta-\tau) \Psi_4\notag \\
& -2 \nu \Phi_{11}-\bar{\nu} \Phi_{20}+2 \lambda \Phi_{12}  +2(\gamma+\bar{\mu}) \Phi_{21}+(\bar{\tau}-2 \bar{\beta}-2 \alpha) \Phi_{22}, \\
D \Phi_{11}-\delta \Phi_{10}+\Delta \Phi_{00}-  \bar{\delta} \Phi_{01}+3 D \Lambda& =(2 \gamma+2 \bar{\gamma}-\mu-\bar{\mu}) \Phi_{00} +(\pi-2 \alpha-2 \bar{\tau}) \Phi_{01} \notag\\
& +(\bar{\pi}-2 \bar{\alpha}-2 \tau) \Phi_{10}+2(\rho+\bar{\rho}) \Phi_{11} +\bar{\sigma} \Phi_{02}+\sigma \Phi_{20}-\bar{\kappa} \Phi_{12}-\kappa \Phi_{21},\\
D \Phi_{12}-\delta \Phi_{11}+\Delta \Phi_{01}-\bar{\delta} \Phi_{02}+  3 \delta \Lambda& =(2 \gamma-\mu-2 \bar{\mu}) \Phi_{01}+\bar{\nu} \Phi_{00}-\bar{\lambda} \Phi_{10}  +2(\bar{\pi}-\tau) \Phi_{11} +(\pi+2 \bar{\beta}-2 \alpha-\bar{\tau}) \Phi_{02} \notag\\
& +(2 \rho+\bar{\rho}-2 \bar{\epsilon}) \Phi_{12}+\sigma \Phi_{21}-\kappa \Phi_{22},\\
D \Phi_{22}-\delta \Phi_{21}+\Delta \Phi_{11}-\bar{\delta} \Phi_{12}+  3 \Delta \Lambda & =\nu \Phi_{01}+\bar{\nu} \Phi_{10}-2(\mu+\bar{\mu}) \Phi_{11}-\lambda \Phi_{02} -\bar{\lambda} \Phi_{20}+(2 \pi-\bar{\tau}+2 \bar{\beta}) \Phi_{12} \notag\\
& +(2 \beta-\tau+2 \bar{\pi}) \Phi_{21}  +(\rho+\bar{\rho}-2 \epsilon-2 \bar{\epsilon}) \Phi_{22}.
\label{NPBianchi}
\end{align}

It is evident that the choice of NP tetrad has certain additional degrees of freedom. That is, given a set of tetrad, there exists a series of transformations that can ensure the orthogonality conditions eq.\ref{OrthCond} of the tetrad. We refer to these as the gauge degrees of freedom in the NP formalism. All these transformations can be classified into three types:
\setcounter{equation}{13}
\renewcommand{\theequation}{A.\arabic{equation}} 
\begin{equation}
\begin{aligned}
\mathrm{I}: & l \rightarrow l, m \rightarrow m+a l, \bar{m} \rightarrow \bar{m}+\bar{a} l , n \rightarrow n+\bar{a} m+a \bar{m}+a \bar{a} l, \\
\mathrm{II}: & n \rightarrow n, m \rightarrow m+b n, \bar{m} \rightarrow \bar{m}+\bar{b} n,  l \rightarrow l+\bar{b} m+b \bar{m}+b \bar{b} n, \\
\mathrm{III}: & l \rightarrow A^{-1} l, n \rightarrow A n, m \rightarrow e^{i \theta} m ,\bar{m} \rightarrow e^{-i \theta} \bar{m},
\end{aligned}
\end{equation}
where $a, b$ are complex functions, and $A, \theta$ are real functions. The transformation relations of NP scalars can be calculated according to their definitions. The results are as follows \cite{ODonnell:2003lqh}. In type I case, spin coefficients transform as
\setcounter{equation}{0}
\renewcommand{\theequation}{A.15.\arabic{equation}} 
\begin{align}
& \nu \mapsto \nu+a \bar{a} \pi+a \lambda+\bar{a} \mu+\bar{a}^2 \tau+\bar{a}^3 a \kappa+\bar{a}^2 a \rho +\bar{a}^3 \sigma+2 \bar{a} \gamma+2 \bar{a}^2 a \epsilon+2 a \bar{a} \alpha+2 \bar{a}^2 \beta +\Delta \bar{a}+a \bar{a} D \bar{a}+a \bar{\delta} \bar{a}+\bar{a} \delta \bar{a}, \\
& \tau \mapsto \tau+\bar{a} \sigma+a \rho+a \bar{a} \kappa, \\
& \gamma \mapsto \gamma+a \bar{a} \epsilon+a \alpha+\bar{a} \beta+\bar{a} \tau+\bar{a}^2 a \kappa+a \bar{a} \rho+\bar{a}^2 \sigma, \\
& \mu \mapsto \mu+a \pi+\bar{a}^2 \sigma+\bar{a}^2 a \kappa+2 \bar{a} \beta+2 a \bar{a} \epsilon+\delta \bar{a}+a D \bar{a}, \\
& \sigma \mapsto \sigma+a \kappa, \\
& \beta \mapsto \beta+a \epsilon+\bar{a} \sigma+a \bar{a} \kappa, \\
& \lambda \mapsto \lambda+\bar{a} \pi+2 \bar{a} \alpha+2 \bar{a}^2 \epsilon+\bar{a}^2 \rho+\bar{a}^3 \kappa+\bar{\delta} \bar{a}+\bar{a} D \bar{a}, \\
& \rho \mapsto \rho+\bar{a} \kappa ,\\
& \alpha \mapsto \alpha+\bar{a} \epsilon+\bar{a} \rho+\bar{a}^2 a \kappa, \\
& \kappa \mapsto \kappa, \\
& \epsilon \mapsto \epsilon+\bar{a} \kappa, \\
& \pi \mapsto \pi+2 \bar{a} \epsilon+\bar{a}^2 a \kappa+D \bar{a} .
\end{align}
Ricci scalars transform as
\setcounter{equation}{0}
\renewcommand{\theequation}{A.16.\arabic{equation}}
\begin{align}
&\Phi_{22} \mapsto  \Phi_{22}+4 a \bar{a} \Phi_{11}+2 a \Phi_{21}+2 \bar{a} \Phi_{12}  +\bar{a}^2 a^2 \Phi_{00}+2 \bar{a} a^2 \Phi_{10}+2 \bar{a}^2 a \Phi_{01}+\bar{a}^2 \Phi_{02}+a^2 \Phi_{20}, \\
&\Phi_{11} \mapsto  \Phi_{11}+a \bar{a} \Phi_{00}+a \Phi_{10}+\bar{a} \Phi_{01}, \\
&\Phi_{21} \mapsto  \Phi_{21}+2 a \bar{a} \Phi_{10}+\bar{a}^2 \Phi_{01}+a \Phi_{20}+2 \bar{a} \Phi_{11}+\bar{a}^2 a \Phi_{00},\\
&\Phi_{10} \mapsto \Phi_{10}+\bar{a} \Phi_{00},\\
&\Phi_{12} \mapsto \Phi_{12}+2 a \bar{a} \Phi_{01}+a^2 \Phi_{10}+\bar{a} \Phi_{02}+2 a \Phi_{11}+a^2 \bar{a} \Phi_{00},\\
&\Phi_{01} \mapsto \Phi_{01}+a \Phi_{00},\\
&\Phi_{20} \mapsto \Phi_{20}+2 \bar{a} \Phi_{10}+\bar{a}^2 \Phi_{00},\\
&\Phi_{00} \mapsto \Phi_{00},\\
&\Phi_{02} \mapsto \Phi_{02}+2 a \Phi_{01}+a^2 \Phi_{00},\\
&\Lambda \mapsto \Lambda.
\end{align}
Weyl scalars transform as
\setcounter{equation}{0}
\renewcommand{\theequation}{A.17.\arabic{equation}}
\begin{align}
& \Psi_0 \mapsto \Psi_0, \\
& \Psi_1 \mapsto \Psi_1+\bar{a} \Psi_0, \\
& \Psi_2 \mapsto \Psi_2+2 \bar{a} \Psi_1+\bar{a}^2 \Psi_0, \\
& \Psi_3 \mapsto \Psi_3+3 \bar{a} \Psi_2+3 \bar{a}^2 \Psi_1+\bar{a}^3 \Psi_0 ,\\
& \Psi_4 \mapsto \Psi_4+4 \bar{a} \Psi_3+6 \bar{a}^2 \Psi_2+4 \bar{a}^3 \Psi_1+\bar{a}^4 \Psi_0.
\end{align}

In type II case, spin coefficients transform as
\setcounter{equation}{0}
\renewcommand{\theequation}{A.18.\arabic{equation}}
\begin{align}
\kappa \mapsto & \kappa+b \bar{b} \tau+\bar{b} \sigma+b \rho+b^2 \pi+b^3 \bar{b} \nu+b^2 \bar{b} \mu+b^3 \lambda+b \epsilon+2 b^2 \bar{b} \gamma  +2 b \bar{b} \beta+2 b^2 \alpha-D b-b \bar{b} \Delta b-\bar{b} \delta b-b \bar{\delta} b, \\
\pi \mapsto & \pi+b \lambda+\bar{b} \mu+b \bar{b} \nu, \\
\epsilon \mapsto & \epsilon+b \bar{b} \gamma+\bar{b} \beta+b \alpha+b \pi+b^2 \bar{b} \nu+b \bar{b} \mu+b^2 \lambda, \\
\rho \mapsto & \rho+\bar{b} \tau+b^2 \lambda+b^2 \bar{b} \nu+2 b \alpha+2 b \bar{b} \gamma-\bar{\delta} b-\bar{b} \Delta b, \\
\lambda \mapsto & \lambda+\bar{b} \nu, \\
\alpha \mapsto & \alpha+\bar{b} \tau+b \lambda+b \bar{b} \nu, \\
\sigma \mapsto & \sigma+b \tau+2 b \beta+2 b^2 \gamma+b^2 \mu+b^3 \nu-\delta b-b \Delta b, \\
\mu \mapsto & \mu+b \nu, \\
\beta \mapsto & \beta+b \gamma+b \mu+b^2 \nu, \\
\nu \mapsto & \nu, \\
\gamma \mapsto & \gamma+b \nu, \\
\tau \mapsto & \tau+2 b \gamma+b^2 \nu-\Delta b.
\end{align}
Ricci scalars transform as
\setcounter{equation}{0}
\renewcommand{\theequation}{A.19.\arabic{equation}}
\begin{align}
& \Phi_{00} \mapsto \Phi_{00}+4 b \bar{b} \Phi_{11}+2 \bar{b} \Phi_{01}+2 b \Phi_{10}+b^2 \bar{b}^2 \Phi_{22} +2 b \bar{b}^2 \Phi_{12}+2 b^2 \bar{b} \Phi_{21}+b^2 \Phi_{20}+\bar{b}^2 \Phi_{02}, \\
& \Phi_{11} \mapsto \Phi_{11}+b \bar{b} \Phi_{22}+\bar{b} \Phi_{12}+b \Phi_{21}, \\
& \Phi_{01} \mapsto \Phi_{01}+2 b \bar{b} \Phi_{12}+b^2 \Phi_{21}+\bar{b} \Phi_{02}+2 b \Phi_{11}+b^2 \bar{b} \Phi_{22}, \\
& \Phi_{12} \mapsto \Phi_{12}+b \Phi_{22}, \\
& \Phi_{10} \mapsto \Phi_{10}+2 b \bar{b} \Phi_{21}+\bar{b} \Phi_{12}+b \Phi_{20}+2 \bar{b} \Phi_{11}+\bar{b}^2 b \Phi_{22}, \\
& \Phi_{21} \mapsto \Phi_{21}+\bar{b} \Phi_{22}, \\
& \Phi_{02} \mapsto \Phi_{02}+2 b \Phi_{12}+b^2 \Phi_{00}, \\
& \Phi_{22} \mapsto \Phi_{22}, \\
& \Phi_{20} \mapsto \Phi_{20}+2 \bar{b} \Phi_{21}+\bar{b}^2 \Phi_{22}, \\
& \Lambda \mapsto \Lambda.
\end{align}
Weyl scalars transform as
\setcounter{equation}{0}
\renewcommand{\theequation}{A.20.\arabic{equation}}
\begin{align}
& \Psi_0 \mapsto \Psi_0+4 b \Psi_1+6 b^2 \Psi_2+4 b^3 \Psi_3+b^4 \Psi_4, \\
& \Psi_1 \mapsto \Psi_1+3 b \Psi_2+3 b^2 \Psi_3+b^3 \Psi_4, \\
& \Psi_2 \mapsto \Psi_2+2 b \Psi_3+b^2 \Psi_4,\\
& \Psi_3 \mapsto \Psi_3+b \Psi_4, \\
& \Psi_4 \mapsto \Psi_4 .
\end{align}

In type III case, spin coefficients transform as
\setcounter{equation}{0}
\renewcommand{\theequation}{A.21.\arabic{equation}}
\begin{align}
\kappa & \mapsto A^2 e^{i \theta} \kappa, \\
\pi & \mapsto e^{-i \theta} \pi, \\
\epsilon & \mapsto \frac{A}{2}\left[\frac{1}{A} D A+e^{-i \theta} D e^{i \theta}+2 \epsilon\right], \\
\rho & \mapsto A \rho, \\
\lambda & \mapsto \frac{1}{A} e^{-2 i \theta} \lambda, \\
\alpha & \mapsto \frac{1}{2} e^{-i \theta}\left[\frac{1}{A} \bar{\delta} A+e^{-i \theta} \bar{\delta} e^{i \theta}+2 \alpha\right] ,\\
\sigma & \mapsto A e^{2 i \theta} \sigma ,\\
\mu & \mapsto \frac{1}{A} \mu, \\
\beta & \mapsto \frac{1}{2} e^{-i \theta}\left[\frac{1}{A} \delta A+e^{-i \theta} \delta e^{i \theta}+2 \beta\right], \\
\nu & \mapsto \frac{1}{A^2} e^{-i \theta} \nu ,\\
\gamma & \mapsto \frac{1}{2 A}\left[\frac{1}{A} \Delta A+e^{-i \theta} \Delta e^{i \theta}+2 \gamma\right], \\
\tau & \mapsto e^{i \theta} \tau .
\end{align}
Ricci scalars transform as
\setcounter{equation}{0}
\renewcommand{\theequation}{A.22.\arabic{equation}}
\begin{align}
& \Phi_{00} \mapsto A^2 \Phi_{00}, \\
& \Phi_{11} \mapsto \Phi_{11}, \\
& \Phi_{01} \mapsto A e^{i \theta} \Phi_{01}, \\
& \Phi_{12} \mapsto A^{-1} e^{i \theta} \Phi_{12}, \\
& \Phi_{10} \mapsto A e^{-i \theta} \Phi_{10}, \\
& \Phi_{21} \mapsto A^{-1} e^{-i \theta} \Phi_{21}, \\
& \Phi_{02} \mapsto e^{2 i \theta} \Phi_{02}, \\
& \Phi_{22} \mapsto A^{-2} \Phi_{22}, \\
& \Phi_{20} \mapsto e^{-2 i \theta} \Phi_{20},\\
&\Lambda \mapsto \Lambda.
\end{align}
Weyl scalars transform as
\setcounter{equation}{0}
\renewcommand{\theequation}{A.23.\arabic{equation}}
\begin{align}
& \Psi_0 \mapsto A^2 e^{2 i \theta} \Psi_0, \\
& \Psi_1 \mapsto A e^{i \theta} \Psi_1, \\
& \Psi_2 \mapsto \Psi_2, \\
& \Psi_3 \mapsto A^{-1} e^{-i \theta} \Psi_3, \\
& \Psi_4 \mapsto A^{-2} e^{-2 i \theta} \Psi_4.
\end{align}

There also exists some discrete transformations which keep the full set of NP equations invariant. One of them is prime operation
\begin{equation}
l^{\mu} \mapsto l'^{\mu}= n^{\mu}, \quad n^{\mu} \mapsto n'^{\mu}= l^{\mu}, \quad m^{\mu} \mapsto m'^{\mu}= \bar{m}^{\mu}, \quad \bar{m}^{\mu} \mapsto \bar{m}'^{\mu}= m^{\mu}.
\end{equation}
And spin coefficients transform as
\begin{equation}
\begin{aligned}
\kappa^{\prime} = -\nu, \quad \pi^{\prime} = -\tau, \quad \epsilon^{\prime} = -\gamma, \quad \rho^{\prime} = -\mu, \quad \lambda^{\prime} = -\sigma, \quad \alpha^{\prime} = -\beta, \\
\sigma^{\prime} = -\lambda, \quad \mu^{\prime} = -\rho, \quad \beta^{\prime} = -\alpha, \quad \nu^{\prime} = -\kappa, \quad \gamma^{\prime} = -\epsilon, \quad \tau^{\prime} = -\pi.
\end{aligned}
\end{equation}
Ricci scalars transform as
\begin{equation}
\begin{aligned}
& \Phi_{00}^{\prime} = \Phi_{22}, \quad \Phi_{11}^{\prime} = \Phi_{11}, \quad \Phi_{22}^{\prime} = \Phi_{00}, \quad \Phi_{10}^{\prime} = \Phi_{12}, \quad \Phi_{20}^{\prime} = \Phi_{02}, \\
& \Phi_{12}^{\prime} = \Phi_{10}, \quad \Phi_{01}^{\prime} = \Phi_{21}, \quad \Phi_{02}^{\prime} = \Phi_{20}, \quad \Phi_{21}^{\prime} = \Phi_{01}, \quad \Lambda^{\prime} = \Lambda.
\end{aligned}
\end{equation}
Weyl scalars transform as
\begin{equation}
\Psi_0^{\prime} = \Psi_4, \quad \Psi_1^{\prime} = \Psi_3, \quad \Psi_2^{\prime} = \Psi_2, \quad \Psi_3^{\prime} = \Psi_1, \quad \Psi_4^{\prime} = \Psi_0.
\end{equation}

\end{appendices}
\bibliographystyle{apj}
\bibliographystyle{apsrev4-1}
\bibliography{ModifiedTeukolsky}

\end{document}